\documentstyle[12pt]{article}
\input{epsf}
\def\And{{\rm and\ }}

\def\arctg{\mathop{\rm arctg}\nolimits}

\def\stars{\bigskip\centerline{***}\medskip}

\newif\ifboo \boofalse

\def\Review#1{\boofalse{\it #1},}
\def\Name#1{{\sc #1},}
\def\Vol#1{\ifboo Vol. {\bf #1}\else{\bf #1}\fi}
\def\Year#1{\ifboo #1\else(#1)\fi}
\def\Book#1{\bootrue{\it #1},}
\def\Page#1{\ifboo {\rm p. #1}\else{\rm #1}\fi}

\begin{document}
%
%
%
\title{Orientational transition in nematic liquid crystals under
oscillatory Poiseuille flow}
\author{P. T\'oth$^{1}$, A. P. Krekhov$^{1,2}$, L. Kramer$^{1}$ 
and J. Peinke$^{3}$
\\ \\
$^{1}$ Physikalisches Institut, Universit\"at Bayreuth,\\
95440 Bayreuth, Germany\\
$^{2}$ Institute of Molecule and Crystal Physics,\\
Russian Academy of Sciences, 450025 Ufa, Russia\\
$^{3}$ Fachbereich Physik, Universit\"at Oldenburg,\\
26111 Oldenburg, Germany}
%
%
%
%
%
%
%
\date{}
\maketitle
\begin{abstract}
We investigate the orientational behaviour of a homeotropically aligned
nematic liquid crystal subjected to an oscillatory plane Poiseuille
flow produced by an alternating pressure gradient.
For small pressure amplitudes the director oscillates within the flow
plane around the initial homeotropic position, whereas for higher
amplitudes a spatially homogeneous transition to out-of-plane
director motion was observed for the first time.
The orientational transition was found to be supercritical and the
measured frequency dependence of the critical pressure amplitude in the 
range between $2$ and $20$ Hz was in quantitative agreement with
a recent theory.
\end{abstract}

\section{Introduction}
In nematic liquid crystals (NLCs) complex behaviour arises 
from the strong coupling between velocity ${\bf v}$ and director
${\bf \hat n}$.
For a steady flow along the $x$ axis with a velocity field $v_x(z)$,
$v_y=v_z=0$ (rectilinear plane flow, typically of the Couette or 
Poiseuille type) the director will, in the absence of other torques,
align in the flow plane ($x-z$ plane) at the angle
$\theta_{fl}=\pm\arctg(\sqrt{\alpha_3/\alpha_2})$
with the $x$ axis if $\alpha_3/\alpha_2 > 0$ (the $\pm$ sign corresponds
to positive/negative shear rate $\partial v_x/\partial z$)
\cite{Leslie68,Pikin73}.
In common nematics $\alpha_3/\alpha_2$ is small but positive
($\approx 0.01$), however, in some materials (in particular near a
nematic - smectic transition) one has $\alpha_3/\alpha_2 < 0$ and
instead of flow alignment there is more complicated tumbling motion
\cite{CT75,PG76,PGP76,Manneville81,ZL89ab}.
Note that $\alpha_2$ is always negative, so the sign of
$\alpha_3/\alpha_2$ is determined by that of $\alpha_3$.

When the velocity field oscillates periodically and symmetrically around
zero [time average $\langle v_x(z,t) \rangle = 0$] the situation becomes
more complicated.
Here we are concerned with {\it oscillatory Poiseuille flow} where the
velocity field $v_x(z,t)$ is induced by applying an alternating pressure
difference $\Delta P \cos(\omega t)$ between the ends of a plane 
cell filled by a NLC (bounding plates parallel to the $x-y$ plane).
The planarly aligned case with the director oriented perpendicular to
the flow plane, {\it i.e.\/}, in the $y$ direction, has been studied
intensively
in experiments by Pieranski and Guyon \cite{GP75,JPG76} and
theoretically by Dubois-Violette and Manneville
\cite{MD-V76,Manneville79}.
For very low frequencies a spatially homogeneous instability at some 
critical pressure amplitude has been found whereas for higher 
frequencies a transition to periodic rolls oriented along the 
flow direction was observed \cite{GP75,JPG76}.

When the director is prealigned within the flow plane the
behaviour under oscillatory Poiseuille flow can be quite complex and
has not yet been clarified completely.
In the high-Ericksen number limit where elasticities (and therefore also
boundaries) are neglected the time-averaged torque is zero and one has
reversible behaviour (no attractors).
Thus one expects the existence of a two-parameter family of solutions
for the director oscillations depending on the initial orientation.
When the effect of elasticities, which represent singular 
perturbations, is included {\it and} the shear rate is spatially 
nonhomogeneous (as is the case for Poiseuille, but not for Couette 
flow), there arizes an averaged torque, which produces attractors
\cite{KK94}.
It is then found that, without surface anchoring, the flow-alignment
solution is always unstable and one is left with one major attractor, 
which for $\alpha_3/\alpha_2 < 0$ is
the orientation perpendicular to the flow plane, or, for 
$\alpha_3/\alpha_2 > 0$, oscillations around an orientation close to it
(in the $y-z$ plane).
In this case also the perpendicular orientation becomes unstable which
is consistent with the results of \cite{GP75,JPG76,MD-V76,
Manneville79}.
In both cases, the oscillations in the flow plane around the homeotropic
direction are unstable.
Including boundary effects one in fact expects a threshold for an 
out-of-pane transition \cite{KK94}.

In this letter we give the first experimental confirmation of this
effect.
We also present new, rigorous numerical calculations of the threshold
as a function of frequency and compare with the experimental results.

\section{Experimental setup}
Experiments were carried out with the nematic MBBA
[N-(4-methoxy\-benzylidene)-4-butyl\-aniline], a flow aligning
material ($\alpha_3$, $\alpha_2 < 0$).
The nematic was confined between two $3$ mm thick glass plates
of lateral dimensions $L_x=15$ mm $\times$ $L_y=20$ mm.
The thickness $d$ of the nematic layer was controlled by two 
polyester-foil spacers placed along the shorter sides at $y=\pm L_y/2$
(nominal thickness $23$ $\mu$m).
They also served to close the flow cell along these sides.
The other sides were open.
The nematic was held inside the cell by capillary forces.
To induce homeotropic alignment of the director at the glass plates
the inner surfaces were treated with the surfactant
DMOAP [dimethyloctadecyl-(3-trimethoxysilyl)-propylammonium chloride].

\begin{figure}
\begin{center}
\epsfxsize=12cm
\vspace*{-1.0cm}
\epsfbox{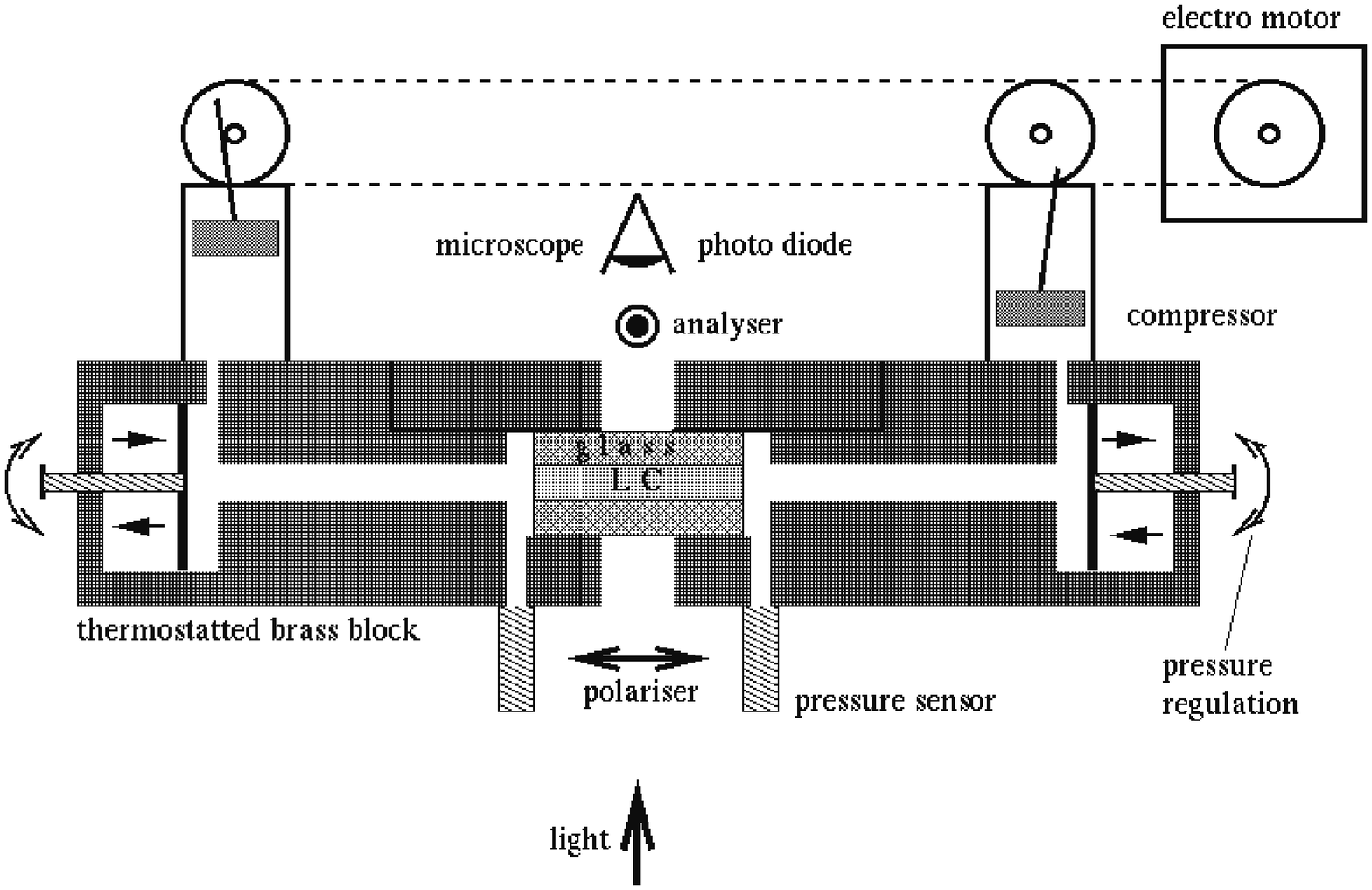}
\end{center}
\caption{Cross section of the experimental setup.}
\label{fig1}
\end{figure}

The setup surrounding the cell was designed to apply an
oscillatory air pressure difference
$P_{1,2} = \Delta P_{1,2} \cos(\omega t +\varphi_{1,2})$ to
the two open sides of the cell (fig. \ref{fig1}).
The cell was mounted inside a brass block
of outer dimensions $4 \times 5 \times 10$ cm.
The open sides of the cell were in contact with channels in the brass
block which widened towards the cell in order to gain the width $L_y$.
As a result the pressure was distributed rather evenly over the entire 
width of the cell.
The channels connected, via cavities of adjustable volume,
to two vertical cylinders accommodating pistons
driven sinusoidally in opposing sense by an electromotor.
The displacement was $v_d=5$ cm$^3$ and the frequency $f=\omega/(2\pi)$
ranged from $1$ to $20$ Hz.
The volume of the  cavities could be adjusted by screws in order to
regulate the ``passive'' volume $V_i$ of air between the lower position
of the pistons and the cell.
This allowed to vary the amplitude $\Delta P$ of the applied pressure in
a range from $1$ to $20$ kPa.
Pressure sensors were connected to the open sides of the cell
in order to measure the pressure head with respect
to the surrounding air.

The response of the nematic was detected optically by exploiting
its birefringence.
We used a light source with maximum intensity at wavelength $610$ nm
(cold source with filter) and  crossed polars (polariser 
$\parallel \hat{e}_x$, analyser $\parallel \hat{e}_y$).
The transmitted light intensity was measured by a semiconductor detector,
either integrating over an area of $28.3$ mm$^2$
or with a spatial resolution of $2$ $\mu$m (long-range microscope).
The photo detector in the microscope could be replaced by a CCD camera.
Interfaces for the signal processing were either an A/D board or
a frame grabber with a resolution of $512 \times 512$ pixels and
$256$ grey scales.
The temperature of the brass block was kept at $25.0 \pm 0.1 ^\circ$C
by a water thermostat.

Due to the large aspect ratio of the flow cell ($L_y/d \approx 10^3$)
one may expect rectilinear Poiseuille flow ($v_y=v_z=0$) except in thin
boundary layers near the edges.
In order to test the flow field and to calibrate
the pressure sensors the motion of Al$_2$O$_3$ tracers
(diameter $5$ $\mu$m) was recorded with the CCD camera and analysed.
No $y$ component of the flow was detected.

An alternating pressure applied to a fluid layer decays by compressibility
effects over a length
$\delta=d\sqrt{2\rho c^2/(\pi^2\omega\eta)}$ \cite{LL86,BDRSY90}.
Here $\rho$ is the mass density and $c$ is the speed of
sound. The viscosity for the homeotropic orientation is given by
$\eta=(-\alpha_2+\alpha_4+\alpha_5)/2$.
Using MBBA material parameters ($\rho=10^3$ kg/m$^3$, $c=1.5 \cdot 10^3$
m/s \cite{EGW73}, $\eta=135.4 \cdot 10^{-3}$ N$\cdot$s/m$^2$), cell
thickness $d=23$ $\mu$m, and the frequency $f=20$ Hz one obtains
$\delta\approx 12$ cm which is considerably larger than the cell
length $L_x=15$ mm.
Moreover, since in our setup the pressure difference is applied
symmetrically at both sides of the capillary the flow is expected to be
practically homogeneous along the $x$ direction.
This was also verified by tracing the Al$_2$O$_3$ particles.

\section{Experimental results and comparison with theory}
\begin{figure}
\begin{center}
\epsfxsize=12cm
\epsfbox{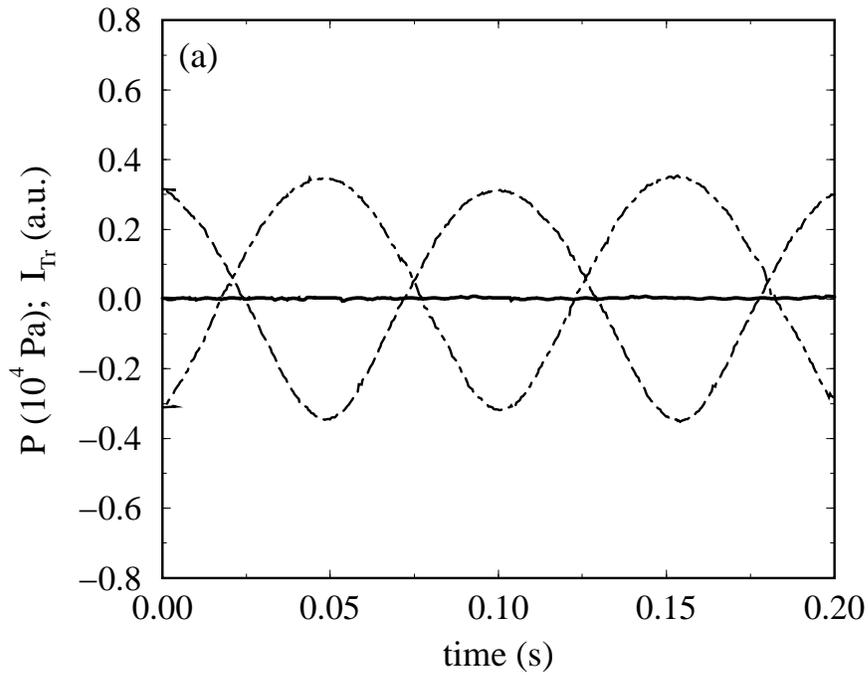}
\epsfxsize=12cm
\epsfbox{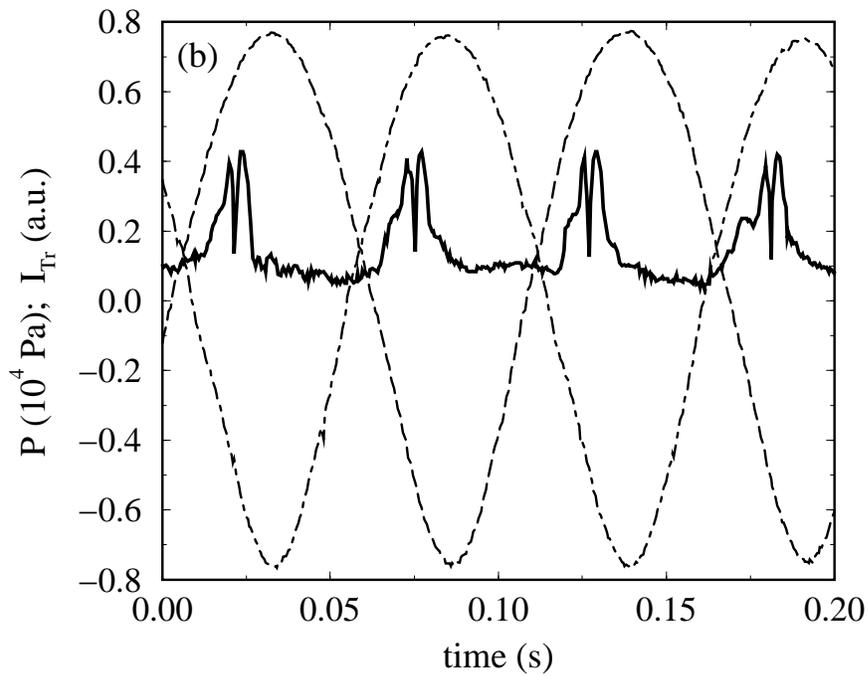}
\end{center}
\caption{Signal of the pressure sensors
(dashed and dot-dashed lines) and transmitted light
intensity (solid line) for a subcritical pressure (a) and 
a supercritical pressure (b).}
\label{fig2}
\end{figure}

Let us define the director components as 
${\bf \hat n}=(\sin\theta \cos\phi , \sin\theta \sin\phi , cos\theta)$
with the polar angle $\theta$ measured with respect to the $z$ axis and 
azimuthal angle $\phi$ measured from the $x$ axis.
For amplitudes $\Delta P$ below any instability the director is expected
to oscillate within the flow plane, {\it i.e.\/} $\phi=0$, and 
in our geometry of crossed polars the intensity should be minimal
(and constant in time).
As a result of an out-of-plane transition the intensity should increase
except when  $\phi = \pi/2$ and $\phi = \pi$.
In fig. \ref{fig2} we show examples of the signals of the two pressure 
sensors
together with the signals of the transmitted light intensity measured 
by the
photo diode as functions of time for an amplitude below and above 
threshold.
We have $\varphi_2=\varphi_1 + \pi$ and 
$\Delta P_1 = \Delta P_2 = \Delta P/2$.

The signal of the light intensity in fig. \ref{fig2}b exhibits two 
rather sharp double peaks over the oscillatory flow period.
We interpret the minimum as the (rapid) passage of the director through
$\phi=\pi/2$. 
The long low-intensity interval between the peaks then
results from $\phi$ approaching $0$ and $\pi$ while the director 
reverses its motion (trajectories of the director motion are shown in
ref.\cite{KK94}).
Slightly above the out-of-plane instability threshold one obtains for
the azimuthal angle $\phi$ the following expression \cite{KK94}
\begin{equation}
\label{phi_approx}
\tan\phi = \chi/[8 a (z/d) \sin\omega t] ,
\end{equation}
where $\chi$ ($\ll 1$) is the out-of-plane angle between the director 
and
flow plane ($x-z$ plane) when the director passes through $\phi=\pi/2$
and
$a=A/d$ is the dimensionless flow displacement amplitude at midplane.
Clearly $|\phi|$ is maximal at the cell boundaries $z=\pm d/2$ and the
leading contribution to the transmitted light intensity comes from there
(for $\chi \ll 1$).
This gives an approximate time dependence of the transmitted light
intensity
\begin{equation}
\label{Itr_approx}
I_{Tr} \sim \sin^2[2\phi(z=d/2)] 
\sim \left[ \frac{\chi 4 a \sin\omega t}{\chi^2 + 
(4 a \sin\omega t)^2} \right]^2 .
\end{equation}
Equation (\ref{Itr_approx}) describes the observed
signal of the light intensity (fig. \ref{fig2}b) and agrees with the
results of full simulations of the optical response based on the 
numerical solution of Maxwell's equations with the director 
distribution calculated from the nematodynamic equations (see below).
In principle such type of signal could also be caused by an 
elliptical motion of the director without breaking the left-right
symmetry.
We have excluded this possibility by measurements with the analyser
rotating with the driving frequency.

\begin{figure}
\begin{center}
\epsfxsize=12cm
\epsfbox{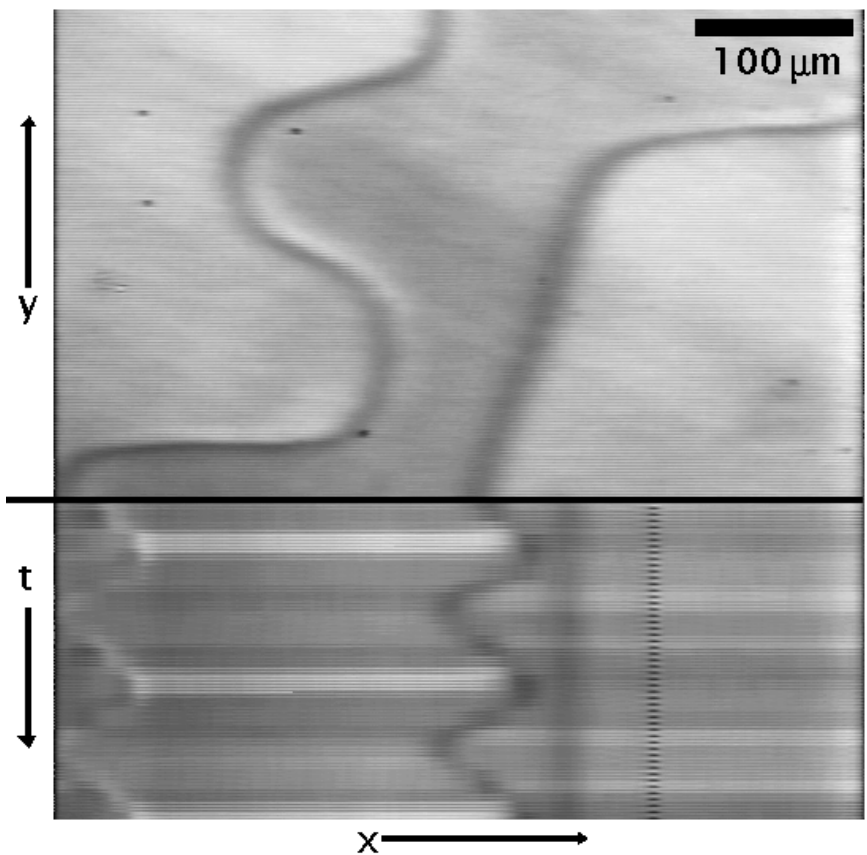}
\end{center}
\caption{Snapshot of patches with left- and right-side out-of-plane 
director configurations (upper part) and 
$x-t$ plot showing time oscillations of domain
boundaries (lower part) far from threshold.
The crossed polars were rotated slightly away from the symmetric
orientation to obtain different intensities in the two configurations.
The dark vertical stripes pertain to dust particles on the glass
plates.}
\label{fig3}
\end{figure}

Since the out-of-plane transition involves the breaking of a two-fold 
symmetry 
one expects the appearance of patches of the two states. 
Indeed, under the microscope we always observed domain boundaries 
indicating patches of $\sim 0.1 - 1$ mm$^2$ area (fig. \ref{fig3}).
After some initial coarsening they did not change much over times of
several minutes, except for oscillatory motion of the boundaries with 
the flow field (see the bottom part of fig. \ref{fig3}).

\begin{figure}
\begin{center}
\epsfxsize=12cm
\epsfbox{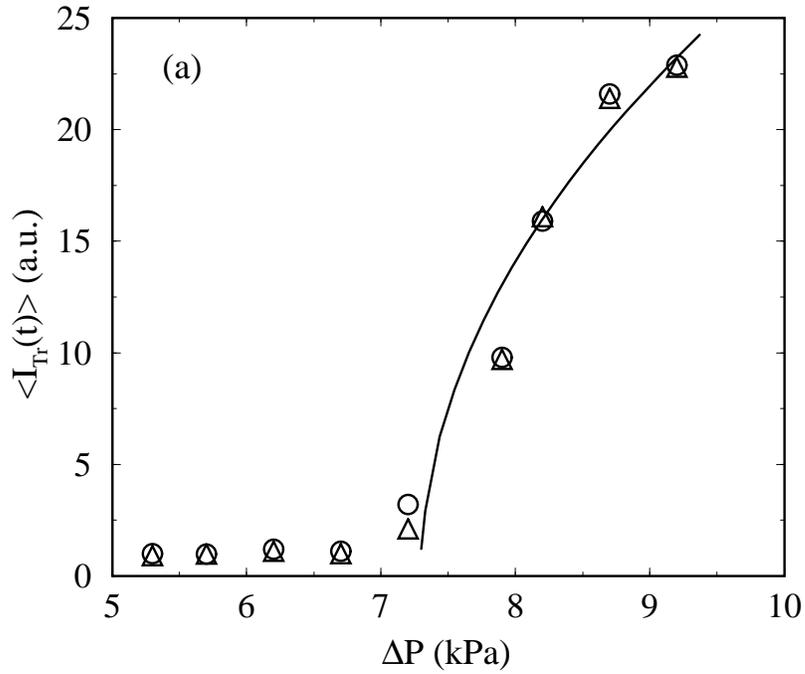}
\epsfxsize=12cm
\epsfbox{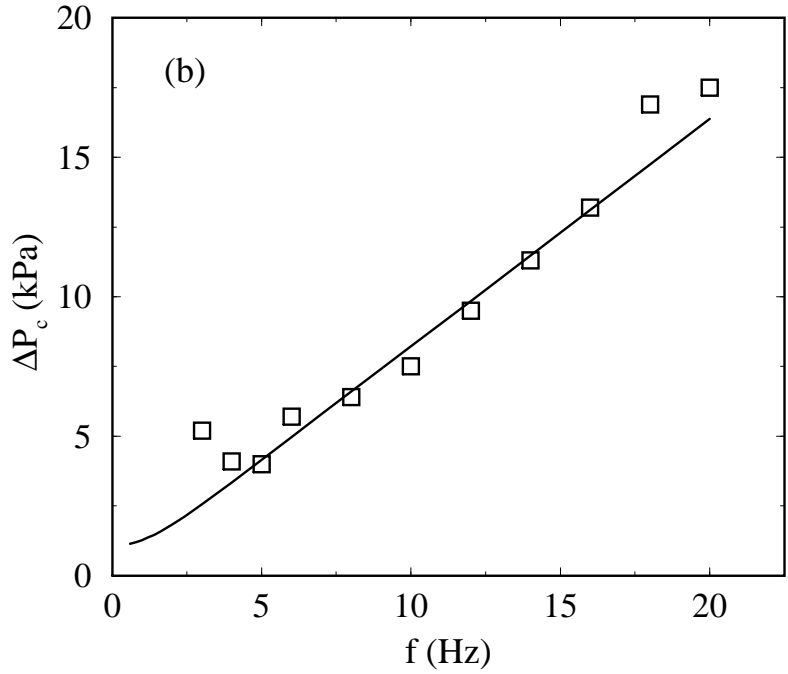}
\end{center}
\caption{a) Transmitted light intensity averaged over the flow period
for increasing ({\Large $\circ$}) and decreasing
({\footnotesize $\triangle$}) pressure amplitude ($f=10$ Hz).
b) Frequency dependence of the critical pressure amplitude 
$\Delta P_c$ for out-of-plane transition: experimental (squares)
and calculated (solid line).}
\label{fig4}
\end{figure}

In order to obtain quantitative results for the threshold we
determined the average intensity of the
transmitted light $\langle I_{Tr}(t) \rangle$.
In fig. \ref{fig4}a $\langle I_{Tr}(t) \rangle$ is shown versus the 
pressure amplitude for a frequency $f=10$ Hz.
By integrating eq.(\ref{Itr_approx}) one finds
$\langle I_{Tr}(t) \rangle \sim \chi$.
Since one expects $\chi \sim \sqrt{\Delta P - \Delta P_c}$ we have
fitted the measured values of $\langle I_{Tr} \rangle$ accordingly.
The best fit curve 
$\langle I_{Tr} \rangle = 16.4 \sqrt{\Delta P-7.3}$ is shown in
fig. \ref{fig4}a (solid).
Clearly the out-of-plane motion sets in at a well-defined threshold 
$\Delta P_c$ via a continuous (forward) bifurcation (no hysteresis),
as predicted by theory \cite{KK94}. 

Direct numerical simulations of the standard set of nematodynamic
equations \cite{GP93} where the director ${\bf \hat n}$ and velocity
${\bf v}$ are functions only of the distance $z$ from the boundaries
and time $t$ (see \cite{KK96} for details) were performed using
finite differences for the spatial derivatives and the
predictor-corrector scheme for the time discretisation.
The critical amplitude of the pressure gradient $\Delta P_c(f)/L_x$ was
calculated for the frequency range $0.5$ Hz $\le f \le$ $20$ Hz,
layer thickness $d=23$ $\mu$m and
MBBA material parameters at $25 ^\circ$C \cite{JCS76,KSS82}
(elastic constants in units of $10^{-12}$ N: $K_{11}=6.66$, 
$K_{22}=4.2$,
and $K_{33}=8.61$; viscosity coefficients in units of $10^{-3}$
N$\cdot$s/m$^2$: $\alpha_1=-18.1$, $\alpha_2=-110.4$, $\alpha_3=-1.1$,
$\alpha_4=82.6$, $\alpha_5=77.9$, $\alpha_6=-33.6$; mass density
$\rho=10^3$ kg/m$^3$).
The results of the numerical computations together with the measured 
values
of $\Delta P_c(f)$ are shown in fig. \ref{fig4}b.
The agreement of the measured values with the calculated ones
is quite good (no adjustable parameters).
Discrepances near the ends of the frequency interval can be explained
by deviations from the sinusoidal driving in these regions.
One notices that for not too low frequencies ($\omega \tau_d \gg 1$,
where $\tau_d$ is the director relaxation time) $\Delta P_c$ rises
linearly with $\omega$.
This corresponds to the critical Poiseuille flow amplitude $a_c$ being
independent of $\omega$, which can be understood from simple scaling
arguments \cite{KK94}.
Quantitative discrepancies with our earlier theoretical work results
from the fact that there the effect of the director distribution on
the flow field, which leads to deviations from the parabolic profile, 
was not included.

\section{Conclusion}
For the first time the out-of-plane orientational transition in
homeotropically oriented nematic liquid crystal under oscillatory
Poiseuille flow was observed experimentally.
The phenomenon is interesting because the underlying torque on the
director resulting from spatially inhomogeneous driving in the
presence of diffusive coupling, arises quite unexpectedly.

The analysis of the transmitted light intensity shows that the 
instability has a supercritical character without hysteresis.
The frequency dependence of the critical pressure amplitude is in a good
quantitative agreement with theoretical results without any adjustable
parameters.
Further efforts will be made to increase the range of
frequency and pressure of the measurements.
The theory predicts that at high driving amplitudes the director
saturates at an orientation perpendicular to the flow plane
($\alpha_3>0$) or near to it ($\alpha_3>0$) \cite{KK94}, unless a
further transition to a spatially periodic state intervenes.

Let us finally mention that a more complicated scenario is expected
for boundary anchoring corresponding to the
flow-alignment angle or, alternatively, simple planar anchoring of the
director  with an additional magnetic field in the flow plane at an
angle of $45^\circ$ with respect to the $x$ axis. 
Then a transition to a slow time-periodic
director precession has been predicted theoretically \cite{KK96}.
This director motion emerges with increasing flow amplitude through a
homoclinic bifurcation and disappears through a Hopf bifurcation.
It would be interesting to perform experiments to test this prediction.

\stars
We thank O. Tarasov for fruitful discussions.
A.K. wishes to thank the University of Bayreuth for their hospitality
and P.T. wishes to acknowledge the hospitality of the Institute of
Molecule and Crystal Physics (Ufa).
Financial support from DFG Grants No. Kr690/12-1, 436RUS113/220, 
INTAS Grant No. 96-498, EU Grant TMR-ERBFM-RXCT960085 and VW Grant
I/72 920 is gratefully acknowledged.
\clearpage
\end{document}